\documentclass[aps,prl,twocolumn,groupedaddress]{revtex4-1}
\usepackage{graphicx}
\usepackage{epsfig}
\begin{document}

\title{Improved Geometrical Scaling at the LHC}
\author{Michal Praszalowicz}
\affiliation{M. Smoluchowski Institute of Physics, Jagellonian University, Reymonta 4,
30-059 Krakow, Poland}

\date{\today}

\begin{abstract}
We show that geometrical scaling exhibited by the $p_{\rm T}$ spectra
measured by the CMS collaboration at the LHC is substantially improved if
the exponent $\lambda$ of the saturation scale depends on $p_{\rm T}$.
This dependence is shown to be the same as the dependence of small $x$
exponent of $F_2$ structure function in deep inelastic scattering taken at
the scale $p_{\rm T}\simeq Q/2$.
\end{abstract}


\maketitle

Recently in Refs.\cite{McLerran:2010wm,McLerran:2010ex} it has been
shown that $p_{\rm T}$ spectra measured by the CMS collaboration
\cite{Khachatryan:2010xs} at the LHC exhibit geometrical scaling.
Geometrical scaling was first introduced in the context of
Golec-Biernat--W\"{u}sthoff (GBW) model \cite{GolecBiernat:1998js} of
deep inelastic scattering (DIS) in Ref.\cite{Stasto:2000er}. There, a reduced
$\gamma^{\ast}$-proton cross-section $\sigma _{\gamma^{\ast}p}(Q^{2},x)$
that in principle depends on two independent kinematical variables:
$Q^{2}$ and Bjorken $x$, for small $x$ (\emph{i.e. }$x<0.01$ or so) does
depend effectively only upon the ratio%
\begin{equation}
\tau=Q^{2}/Q_{\text{sat}}^{2}(x).
\end{equation}
Here $Q_{\text{sat}}(x)$ is so called saturation momentum which is
proportional to the transverse gluon density \cite{GolecBiernat:1998js}%
\begin{equation}
xg(x,Q^{2})\sim\frac{\sigma_{0}}{\alpha_{\text{s}}(Q_{\text{sat}}^{2}%
)}Q_{\text{sat}}^{2}(x) \label{Qg}%
\end{equation}
with $\sigma_{0}$ being dimensional constant that in the GBW model is
equal to 23 mb. Since gluon density rises for small $x$ like a power,
saturation momentum is customarily assumed to take the following form:%
\begin{equation}
Q_{\text{sat}}^{2}(x)=Q_{0}^{2}\left(  x/x_{0}\right)  ^{-\lambda}%
\end{equation}
with constant $\lambda=0.2\div0.3$, determined from the HERA the data.

This new energy scale emerges in the models that incorporate gluon
saturation \cite{Gribov:1984tu,Mueller:1985wy}, like \emph{e.g.} color
glass condensate \cite{McLerran:1993ni,Ayala:1995hx}, although
geometrical scaling itself is more general, and does not require
saturation (\emph{i.e.} gluon density may grow like a power for
arbitrarily small $x$ and the dipole-proton cross-secction needs not to
go to a constant for large dipole sizes). Nevertheless, the existence of
the saturation scale in strong interactions is by now well established.
If so, it should also manifest itself in hadronic collisions, and it
indeed does, as it was shown in
Refs.\cite{McLerran:2010wm,McLerran:2010ex} where a simple Ansatz, based
on dimensional analysis, for the saturation momentum has been proposed:
\begin{equation}
Q_{\text{sat}}^{2}=Q_{0}^{2}\left(  \frac{p_{\text{T}}}{W}\right)  ^{-\lambda}.
\end{equation}
Here $W\sim\sqrt{s}$ and $Q_{0}\sim1$ GeV sets the scale. It turns out
that the charged particle $p_{\text{T}}$ spectra
measured at the LHC at three incident energies: 0.9, 2.36 and 7 TeV exhibit
geometrical scaling, \emph{i.e.} they fall on one energy-independent universal
curve $F(\tau)$
\begin{equation}
\frac{dN_{\text{ch}}}{dydp_{\text{T}}^{2}}=\frac{1}{Q_{0}^{2}\,}\,F(\tau)
\label{GS}
\end{equation}
 if plotted in terms of the scaling variable%
\begin{equation}
\tau=p_{\text{T}}^{2}/Q_{\text{sat}}^{2}%
\end{equation}
with $\lambda\sim0.27$. In essence, geometrical scaling for
the $p_{\text{T}}$ spectra (\ref{GS}) boils down to the prescription
that allows to relate multiplicity
distributions at two different energies $W$ and $W^{\prime}$. If%
\begin{equation}
\frac{dN_{\text{ch}}}{d\eta d^{2}p_{\text{T}}}(p_{\text{T}},W)=\frac{dN_{\text{ch}}%
}{d\eta d^{2}p^{\prime}_{\text{T}}}(p^{\prime}_{\text{T}},W^{\prime}) \label{equality}%
\end{equation}
then the transverse momenta at which Eq.(\ref{equality}) holds, satisfy%
\begin{equation}
p^{\prime}_{\text{T}}=p_{\text{T}}\left(  \frac{W^{\prime}}{W}\right)
^{\lambda/(\lambda+2)}. \label{pTvspT}%
\end{equation}
This formula is independent of $Q_{0}$ and of the overall energy scale of
$W$ or $W^{\prime}$. So the only relevant parameter of geometrical scaling
is exponent $\lambda$.

Equations (\ref{equality}) and (\ref{pTvspT}) allow to {\em rescale}
$p_{\rm T}$ of known spectrum at energy $W$ to another energy
$W^{\prime}$, and {\em predict} $p_{\rm T}$ spectrum at this energy,
provided we know the value of $\lambda$. In the following, transverse
momentum spectra obtained that way will be referred to as {\em rescaled}
spectra.

Alternatively, if we do not know $\lambda$ but we know spectrum at $W$, we
can find $\lambda$ by changing its value until equality (\ref{equality})
is satisfied, {\em i.e.} until the rescaled and true spectra coincide
within errors.

\begin{figure}[h]
\centering
\includegraphics[scale=0.95]{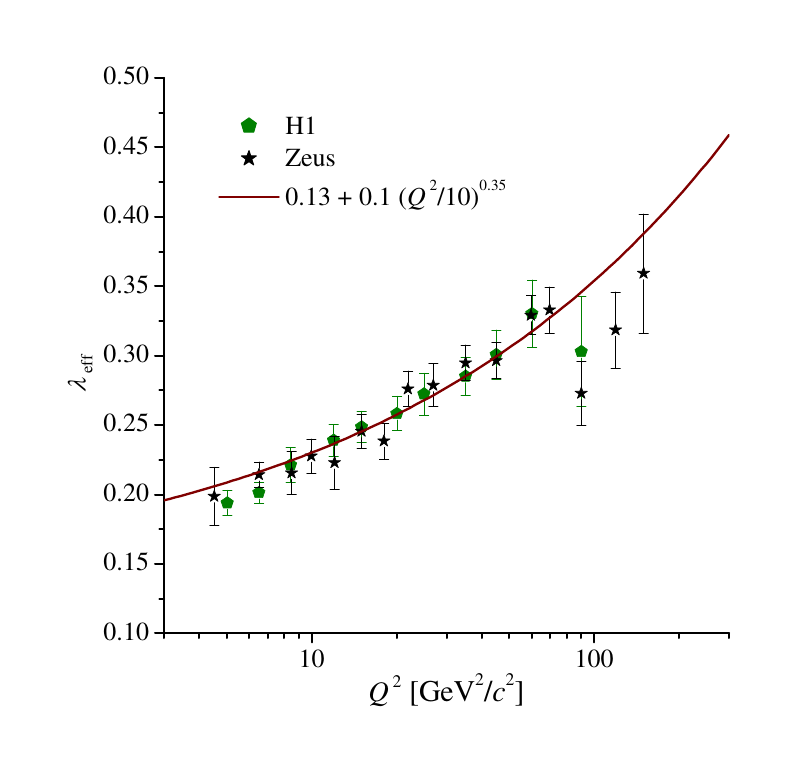}\caption{Dependence of
$\lambda_{\rm eff}$ on $Q^{2}$ from HERA (HERA data points
\cite{HERAdata} after Ref.\cite{Kowalski:2010ue}). }%
\label{lamHERA}%
\end{figure}

In dipole models of DIS the quality of phenomenological fits is further
increased provided one incorporates DGLAP $Q^{2}$ dependence
\cite{Bartels:2002cj} of the saturation scale (\ref{Qg}). Furthermore,
since one "measures" $Q_{\text{sat}}$ with a $Q^2$ dependent probe ({\em
e.g.} with virtual photon or a $p_{\text{T}}$ hadron), effective
saturation scale $Q_{\text{sat,eff}}$ acquires some dependence on the
virtuality of the probe. These two effects may be conveniently accounted
for by replacing $\lambda \rightarrow \lambda (Q^2)$. Indeed, for large
$Q^{2}$ DIS structure function $F_{2}$ that is directly related to the
saturation scale \cite{GolecBiernat:1998js} behaves as:
\begin{equation}
F_{2}(x,Q^{2})\sim\sigma_{0}Q_{\text{sat,eff}}^{2}\sim\frac{1}{x^{\lambda
_{\text{eff}}(Q)}}.%
\end{equation}
Power $\lambda_{\text{eff}}(Q)$ has been extracted from the HERA data
\cite{HERAdata}
(see {\em e.g.} recent Ref.\cite{Kowalski:2010ue} and references therein)
as shown in Fig. \ref{lamHERA}. In the same figure we plot an eyeball fit
to the experimental points given by a simple function%
\begin{equation}
\lambda_{\text{eff}}(Q)=0.13+0.1\left(  \frac{Q^{2}}{10}\right)  ^{0.35}.
\label{lam}%
\end{equation}

An interesting question arises, whether exponent $\lambda$ that governs
geometrical scaling in hadronic collisions exhibits any $p_{\text{T}}$
dependence and, if yes, whether it is similar to the one obtained in DIS.
For $p_{\text{T}}$-dependent $\lambda$ formula (\ref{pTvspT}) takes the
following form:%
\begin{equation}
p_{\text{T}}^{2}\left(  \frac{p_{\text{T}}}{W}\right)
^{\lambda(p_{\text{T}})}=p^{\prime\, 2}_{\text{T}}
\left(  \frac{p^{\prime}_{\text{T}}}{W^{\prime}}\right)
^{\lambda(p^{\prime}_{\text{T}})}. \label{mom12}%
\end{equation}

\begin{figure}[h]
\centering
\includegraphics[scale=0.85]{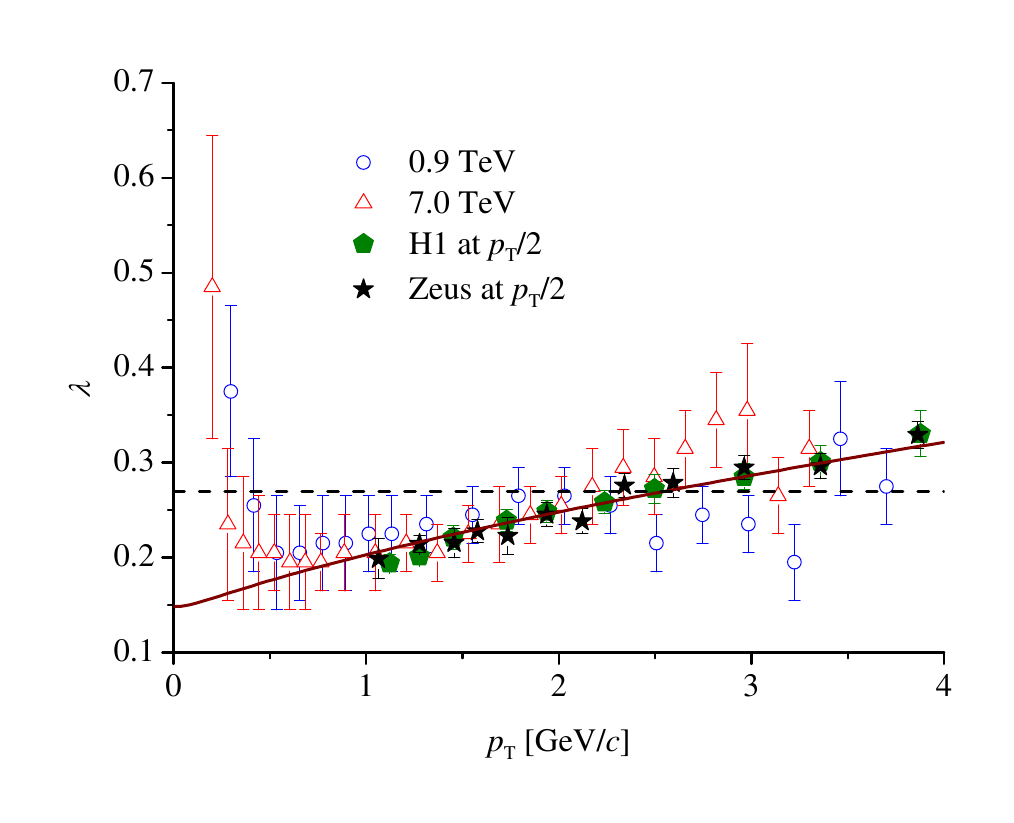}
\caption{Dependence of $\lambda$ on
$p_{\rm T}$. Open circles correspond to $\lambda$
obtained by rescaling 0.9 TeV data to the reference energy
of 2.36 TeV, whereas open triangles correspond to 7 TeV.
Exponent $\lambda_{\rm eff}$ extracted from HERA depicted
by full pentagons (H1) and stars (Zeus) is plotted in function
of $p_{\rm T}=Q/2$ (see text). Solid line corresponds to
Eq.(\ref{lam}) taken at $Q=2 p_{\rm T}$.  }%
\label{lamCMSHERA}%
\end{figure}

\begin{figure}[h]
\centering
\includegraphics[scale=0.70]{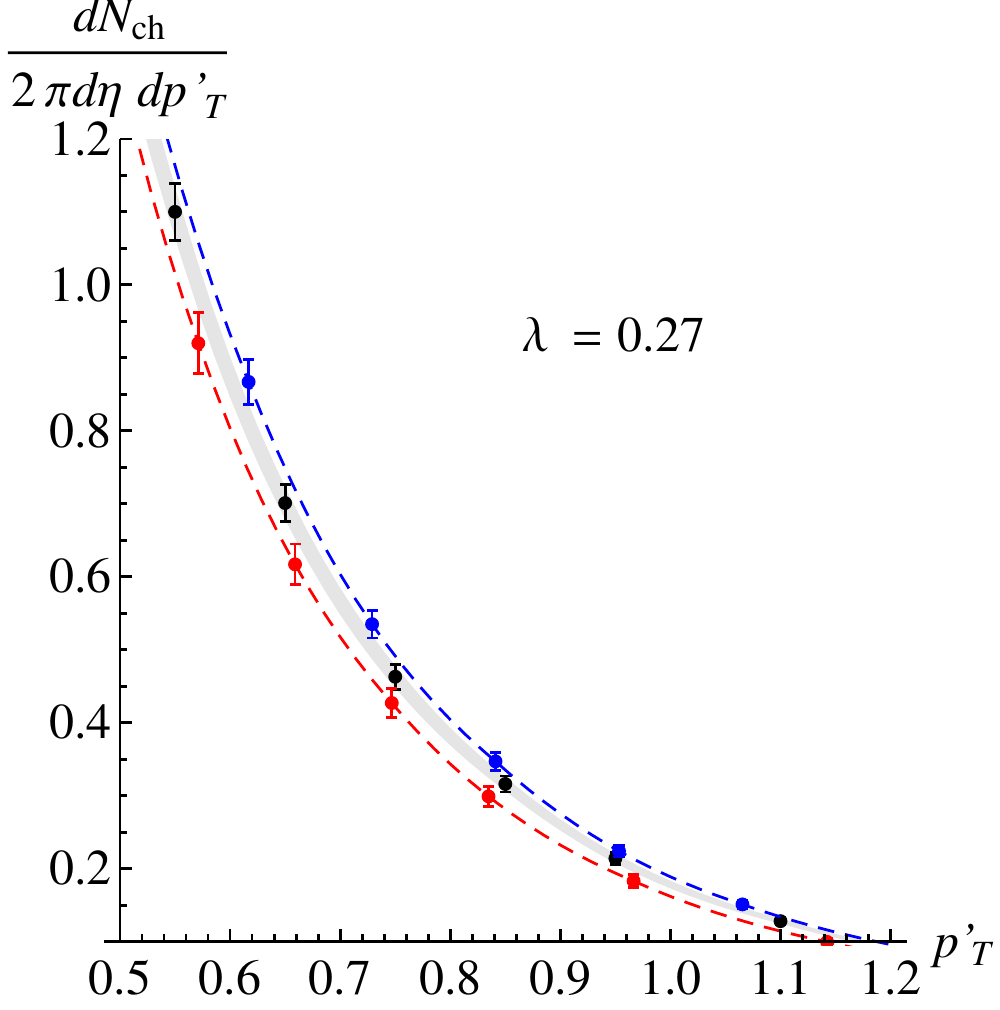}\\
\vspace{0.2cm}
\includegraphics[scale=0.70]{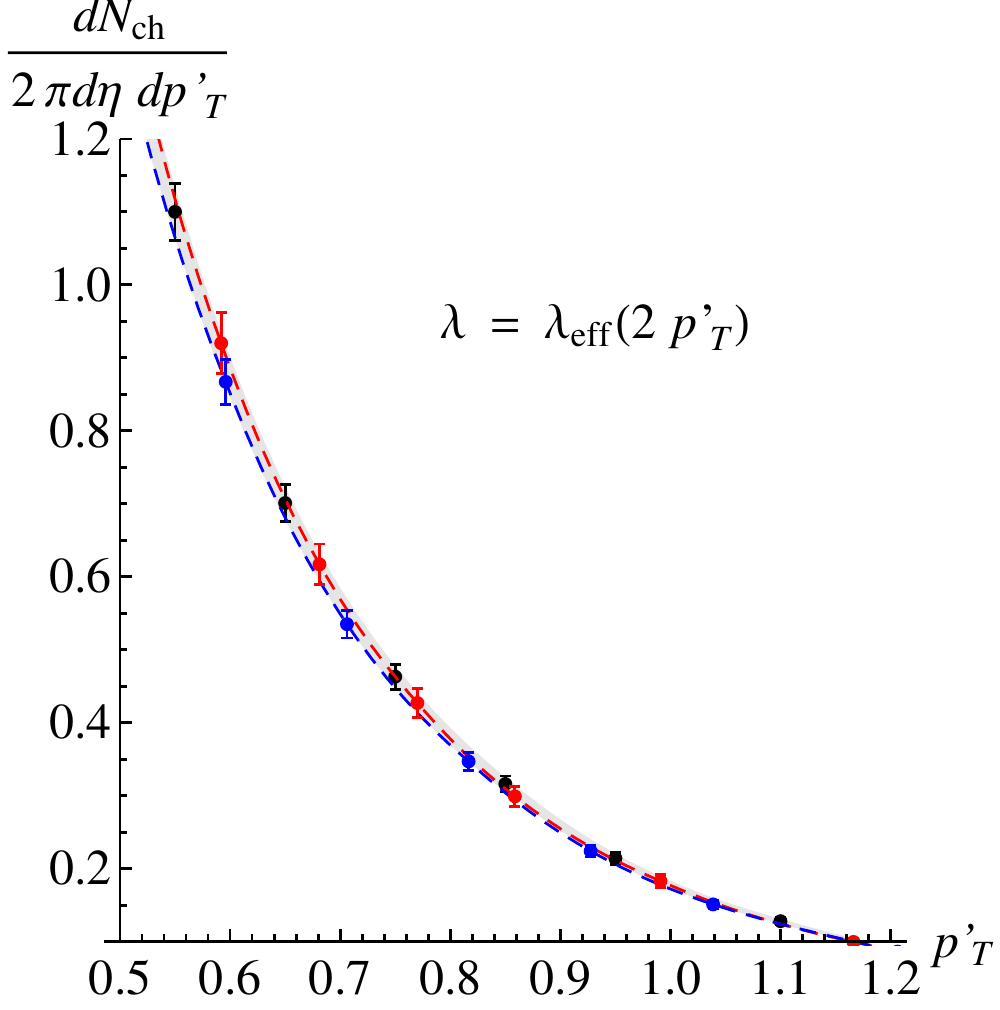}
\caption{Multiplicity
density for $\sqrt{s}=2.36$~TeV (black points and shaded band) as measured by
CMS, and 0.9~TeV (blue points and dashed line) and 7~TeV (red points and
dashed line) spectra rescaled to 2.36~TeV using hypothesis of geometrical scaling
with constant {${\lambda=0.27}$} and "running" effective {$\lambda
_{\mathrm{eff}}(2 p^{\prime}_{\rm T})$} for low {$p_{\rm T}$}.
Horizontal scale in GeV/$c$. }%
\label{to236-1}%
\end{figure}

A simple way to calculate approximate $p_{\text{T}}$ dependence of
$\lambda$ is to use Eq.(\ref{pTvspT}) bin by bin in $p_{\text{T}}$ instead
of an exact equation (\ref{mom12}). For slowly varying
$\lambda(p_{\text{T}})$ such a procedure should give a good first order
approximation. To this end we choose to rescale transverse momenta of CMS
multiplicity spectra at $W=0.9$ and $7$ TeV to the reference energy
$W^{\prime}=2.36$ TeV for some initial value of $\lambda$. Next, we
compare the rescaled spectra with the experimental data at $W^{\prime}$
and repeat the whole procedure until the rescaled and true spectra
coincide (\ref{equality}). In that way we obtain $\lambda(p_{\text{T}})$.
Since in general for $W^{\prime}$ there is no data point at
$p^{\prime}_{\text{T}}$ obtained from (\ref{pTvspT}), we have to
interpolate the reference spectrum and its errors (in the following we
neglect interpolation errors). The result of this interpolation is
depicted in Fig.~\ref{to236-1} by a grey band. In order to estimate the
error of $\lambda(p_{\text{T}})$ we add in quadrature errors of the $W$
spectrum and the interpolated error of the reference $W^{\prime}$ spectrum
at $p_{\rm T}$ and $p^{\prime}_{\rm T}$ respectively. Calling this
effective error $\varepsilon^2$, we repeat the whole procedure solving
equation
\begin{equation}
\frac{dN_{\text{ch}}}{d\eta d^{2}p_{\text{T}}}(p_{\text{T}},W)-\frac{dN_{\text{ch}}%
}{d\eta d^{2}p^{\prime}_{\text{T}}}(p^{\prime}_{\text{T}},W^{\prime})\pm\varepsilon=0
\end{equation}
for $\lambda_{\pm}=\lambda\pm\delta$. The result is plotted in
Fig.\ref{lamCMSHERA}  together with $\lambda_{\text{eff}}$ from DIS taken
at the scale $Q^{2}\simeq 4 p_{\text{T}}^{2}$. Both fit (\ref{lam}) and
the data points are displayed. We see that indeed $\lambda$ does depend on
the transverse momentum. The agreement between $\lambda$ extracted from
the spectrum rescaled from $W=0.9$ TeV (blue circles) and from $W=7$ TeV
(red triangles) is a signature of geometrical scaling. In an interval from
0.5 to approximately 2.5 GeV $\lambda$ rises slowly with increasing
$p_{\text{T}}$. Interestingly, $p_{\rm T}$ dependence of $\lambda$ in this
interval is in a surprising accordance
 with DIS  $\lambda_{\text{eff}}(Q)$ taken at $p_{\text{T}}=Q/2$.
 For higher $p_{\text{T}}$ data become too noisy to draw definite
conclusions. The smooth behavior changes completely for $p_{\text{T}}<0.5$
GeV where the steep rise of $\lambda$ with decreasing $p_{\text{T}}$ is
seen. This may be a signal of an onset of a some other component in the
production mechanism. Here, however, the assumption of slowly varying
$\lambda$ breaks down and more numerical care is needed before
quantitative conclusions concerning small $p_{\text{T}}$ part can be
drawn. One should also stress at this point that further analysis of low
$p_{\text{T}}$ geometrical scaling requires good quality low momentum
data.

Final conclusion that has to be drawn from Fig.\ref{lamCMSHERA} is that
geometrical scaling with constant $\lambda$ is certainly a good first
approximation, but a mild $p_{\text{T}}$ dependence of $\lambda$ improves
substantially the quality of geometrical scaling. This is depicted in
Fig.\ref{to236-1} where we plot the $p_{\text{T}}$ spectra in terms of the
rescaled momentum $p^{\prime}_{\text{T}}$ in the vicinity of 1 GeV where
the difference between constant $\lambda=0.27$ and "running" $\lambda$ of
Eq.(\ref{lam}) is most pronounced. Black points and the shaded band
correspond to the CMS spectrum (and its interpolation) at $2.36$ TeV. Blue
and red points (connected by dashed lines) correspond to 0.9 and 7 TeV
spectra respectively, rescaled to the reference energy of 2.36 TeV for
constant $\lambda$ and "running" $\lambda_{\rm eff}(2 p_{\rm T})$ of
Eq.(\ref{lam}). An improvement for "running" $\lambda$ is evident. For
other $p_{\text{T}}$ intervals where the difference between constant and
"running" $\lambda$ is not large, the quality of geometrical scaling is --
obviously -- comparable.

Geometrical scaling in hadronic collisions is by far less obvious than in
DIS. In DIS we have at our disposal simple theoretical (GBW) model
\cite{GolecBiernat:1998js} that allows to identify kinematical variables
relevant for geometrical scaling. In hadronic collisions such models exist
\cite{Kovchegov:2001sc,Kharzeev:2004if,Kharzeevetal,Levin:2010dw,Tribedy:2010ab}
 but they rely on $k_{\rm T}$ factorization which
has not been proven for soft particle production in central rapidity.
Nevertheless, if $k_{\rm T}$ factorization is assumed, like in the recent
studies of Refs.\cite{Levin:2010dw,Tribedy:2010ab}, then the
proportionality of multiplicity of produced gluons to the saturation
momentum, and therefore geometrical scaling -- assuming local
parton-hadron duality -- can be derived in a rather straightforward way
(see {\em{e.g}} \cite{Kharzeev:2004if}). Nevertheless, the exact form of
the the scaling variable $\tau$, that in principle may depend also on
rapidity, is to some extent a matter of educated guess. Luckily, for
constant $\lambda$ some uncertainties cancel out in Eq.(\ref{pTvspT}),
showing that the only relevant parameter is exponent $\lambda$.

Another notable difference between DIS and hadronic collisions is
that in DIS we deal with totally inclusive cross-section, whereas in pp both
hadronization and final state interactions play essential role. Nevertheless
the imprint of the saturation scale $Q_{\text{sat}}$ is visible in the spectra,
which means that the information on the initial fireball survives until final
hadrons are formed.

In this letter we have shown that the quality of geometrical scaling
improves if the exponent $\lambda$ becomes $p_{\text{T}}$-dependent. We
have computed this dependence by rescaling $p_{\text{T}}$ spectra at 0.9
and 7 TeV to the reference energy 2.36 TeV, however we have also checked
that rescaling 0.9 and 2.36 TeV spectra to 7 TeV or 7 and 2.36 TeV spectra
to 0.9 TeV gives qualitatively the same results. Not only $p_{\text{T}}$
spectra rescaled from different energies  to the  reference energy
$W^{\prime}$ agree (which is the essence of geometrical scaling), but the
$p_{\text{T}}$ dependence of the exponent $\lambda$ agrees with the
dependence obtained from DIS $\lambda_{\rm eff}(Q)$, at the scale
 $Q \sim 2 p_{\text{T}}$. We find this last result remarkable, since
 it provides a direct link between two different types of reactions.

Several points require further clarification. First of all new large
$p_{\text{T}}$ data of good quality will be of importance to test the
range of applicability of geometrical scaling and of the discussed
similarity with DIS. Also low $p_{\text{T}}$ data, where hadronic
$\lambda(p_{\text{T}})$ deviates from the one from DIS, is required to see
whether this deviation signals an onset of a new production mechanism
\emph{common} for different energies, or whether different energies
require different $\lambda(p_{\text{T}})$ violating geometrical scaling in
this region. It will be interesting to verify if geometrical scaling works
also in heavy ion collisions. If so, $p_{\text{T}}$ spectra in heavy ion
collisions measured at different energies and at different centralities
will allow find  $A$ dependence and impact parameter dependence of
$Q_{\text{sat}}$.

\begin{acknowledgments}
The author wants to thank Larry McLerran for a number of stimulating
discussions that triggered this work and Andrzej Bialas for discussion and
careful reading of the manuscript.
\end{acknowledgments}

\end{document}